\documentstyle[aps,twocolumn]{revtex}
\begin{document}
\draft
\preprint{IUCM96-008}
\title{Edge State Transport in Separately Contacted Double-Layer\\
Quantum Hall Systems}
\author{D. Yoshioka\cite{byline} and A.H. MacDonald}
\address{
Department of Physics, Indiana University, Bloomington, IN 47405}
\date{\today}
\maketitle
\begin{abstract}

We develop a theory of edge state transport in separately
contacted double-layer quantum Hall systems which are
tuned close to the resonance condition for tunneling
between the layers.  When applied to the case
where contact is made to only one layer, the theory gives
a quantized Hall resistance and zero longitudinal resistance
in both weak and strong inter-layer coupling limits.
For weak coupling, the leading correction to the Hall resistance
occurs at second order, while the longitudinal resistance
appears at first order.  Recent experiments which show almost
quantized Hall resistance and a substantial longitudinal resistance
are explained and predictions are made for other contact
configurations.

\end{abstract}

\narrowtext

Recently it has become possible to fabricate double-layer two-dimensional
electron systems (2DES's) in which the
two layers can be contacted separately \cite{eisen,patel,ohno}.
This technological achievement has led to the
development of several new experimental techniques for
probing properties of two-dimensional
electronic systems \cite{ohno,eisenexp,pepperexp,sivanexp}.
In one recent experiment Ohno and Sakaki \cite{ohno}
studied transport in the quantum Hall regime for a system in which
contact was made to one layer while the other layer was left floating.
When the system was tuned to equal layer densities to maximize
coupling due to tunneling, they found that the transport
coefficients were usually nearly identical to those measured when both layers
were contacted.  However, near quantum Hall plateaus the Hall
resistivity nearly doubled in the floating layer case,
and the longitudinal resistance rose to a finite value
indicating the presence of dissipation inside the sample.
These authors also observed local maxima in $R_{xx}$
which bracket the local minima at the center of the quantum Hall plateaus.
In this Rapid Communication we report on a theoretical study of transport
in coupled, separately contacted, double-layer systems which was motivated
by these experiments.  We use our theoretical approach to predict
behavior for contact geometries not yet realized experimentally.
\par

We consider a double-layer 2DES
on a quantum Hall plateau and assume that the transport
currents are small enough so that the edge state picture of
the quantum Hall effect is applicable \cite{edgestatetransport}.
For simplicity we assume that a single Landau level is occupied in
each layer and that $k_{B} T$ is small compared to the Landau
level separation $\hbar \omega_{c}$ ; the generalization of our
theory to larger filling factors is obvious.  For a single-layer
system the quantum Hall effect can be understood as a consequence
of the spatial separation between left-going and right-going
states on opposite edges of the sample which permits them
to maintain a local equilibrium even when their chemical
potentials differ and a net current is carried through the sample.
In our model of edge state transport in double-layer systems with the
Hall bar geometry illustrated in Fig.~1,  we assume that equilibration
between left-going states and right-going states is still negligible,
but allow equilibration to occur between the two layers along both
edges.  In the experiments we have in mind, coupling between the
two-layers occurs because of tunneling and we could calculate transport
coefficients for a given contact configuration by using a
multi-probe Landauer-Buttiker formula \cite{butt} and modeling
inter probe transmission coefficients in a way which reflects
the absence of back scattering within an electron layer.
Instead we adopt a slightly different approach which
reflects our belief that phase coherence inside the sample
does not play any essential role in the physics we address.
We emphasize, however, that our discussion could as easily
be presented using the lexicon of Landauer-Buttiker
transport theory, which provides an economical description
of edge state transport in the quantum Hall regime \cite{buttqhe}.

\par
On a quantum Hall plateau, the transport current can be calculated from
the distribution function for edge states.
Our model is based on a semiclassical approach where we
define a local momentum distribution function, $f_{\sigma}(x,p)$,
at each point $x$ along a given edge.
Here $\sigma = \pm$ is used to label the the two layers,
and $p$ is the Landau gauge momentum used to label states
within the lowest Landau level.
The use of a semiclassical distribution function is justified since
the experiments we are interested in are performed on
macroscopic samples.
We assume, as in the Landauer-Buttiker transport theory, that
electrons exiting from a current lead have an equilibrium
thermal distribution with a chemical potential equal to that
of the lead.   The evolution of the distribution function along a
path between current leads is determined by the following
relaxation-time-approximation Boltzmann equation:
\begin{equation}
{{\partial f_{\sigma}(x,p)}\over{\partial x}} =
\frac{1}{\xi} [f_{\sigma}(x,p) - f_{-\sigma}(x,p)]
- \frac{1}{\eta}[f_{\sigma}(x,p) - f_{\sigma}^{T}(x,p)],
\label{boltzman}
\end{equation}
where
\begin{equation}
f_{\sigma}^{T}(x,p) = \{\exp[\epsilon(p)-\mu_{\sigma}(x)]\beta +1\}^{-1}.
\label{fermidistr}
\end{equation}
is the local equilibrium thermal distribution function.
In the two terms on the right hand side of Eq.~(\ref{boltzman})
the phenomenological parameters $\xi$
and $\eta$ are relaxation lengths for inter-edge and
intra-edge equilibration.   This equation describes the evolution
of the distribution function along one edge of the sample; there
is no equilibration between opposite edges on a Hall plateau.
The form of this equation reflects the constant velocity of
edge electrons at the Fermi level which allows the temporal relaxation
to fix the position dependence of the distribution functions.

At each position the distribution function is partially characterized
by a position dependent effective chemical potential $\mu_{\sigma}(x)$,
defined by the following equation:
\begin{equation}
\int_{p_{0}}^{p_{1}} dp [f_{\sigma}(x,p) - f_{\sigma}^{T}(x,p)]
=0.
\label{equilb}
\end{equation}
Here $p_{0}$ and $p_{1}$ are cutoff momenta, which satisfy
$f_{\sigma}(x,p_{1}) \ll 1$ and $1-f_{\sigma}(x,p_{0}) \ll 1$,
so that $\mu_{\sigma}(x)$ is independent of the choice of both $p_{0}$ and
$p_{1}$.
We will see that the only quantity needed for our semiclassical
transport theory is $\mu_{\sigma}(x)$.
In the edge region we may linearize the energy spectrum
letting $\epsilon(p) = v_{F} p$, where $v_{F}$ is the velocity
of electrons in edge states.
With this linearization the Boltzmann equation can be rewritten as a
differential equation for $\mu_{\sigma}(x)$:
\begin{equation}
\frac{d\mu_{\sigma}(x)}{dx} = - \frac{1}{\xi}[\mu_{\sigma}(x)
-\mu_{-\sigma}(x)].
\label{boltchem}
\end{equation}
This equation\cite{note1} is easily solved;
\begin{equation}
\mu_{\sigma}(x) = \frac{1}{2}(1+{\rm e}^{-2x/\xi})\mu_{\sigma}(0)
+ \frac{1}{2}(1-{\rm e}^{-2x/\xi})\mu_{-\sigma}(0).
\label{chemsolv}
\end{equation}
The current along an edge in layer $\sigma$ is given at position $x$ by
\begin{equation}
I_{\sigma}(x) = \frac{e}{h}\int_{p_{0}}^{p_{1}} \frac{d\epsilon(p)}
{dp} f_{\sigma}(x,p).
\label{current}
\end{equation}
In this expression for the current it is important to choose the
interior cutoff $p_{0}$ such that $\epsilon(p_{0})$ has the same value on
right-going and left-going edges of the two layers.
The lower cutoff momentum provides a practical definition of the
edge regions.  (With this definition,
the net contribution to the current from the bulk region
is zero.)
Using the linearization of the energy spectrum
the current carried at position $x$ along an edge
in layer $\sigma$ may also be expressed in terms of
$\mu_{\sigma}(x)$:
\begin{equation}
I_{\sigma} (x) = \frac{e}{h}[\mu_{\sigma}(x) - \epsilon(p_{0})].
\label{currentsolv}
\end{equation}

\par
The contact configurations we wish to model generally include ones in
which voltage probes independently contact a single layer at
some point along one of its edges.  Assuming ideal contacts
in the Landauer-Buttiker transport theory sense, voltage probes
measure the local chemical potential $\mu_{\sigma}(x)$ and although
they alter the distribution function they do not alter the
local effective chemical potential.  Hence they play a
completely  passive role in our theory.  Current probes, on the other
hand,  fix the difference between the incoming current and
outgoing current as well as resetting the chemical potential
of the outgoing electrons.  We apply this theoretical frame work first to
examine the effect on transport in the quantum Hall regime of coupling
to a floating layer.  We apply Eq.~(\ref{chemsolv}) separately to
the left-going and right-going edges as indicated in Fig.~1.
For the right-going edge, the effective chemical potential on the
left side of the contacted ($\sigma = -$) layer
is fixed at the chemical potential of the source ($\mu_{S}$) while
the chemical potential on the left side of the floating layer
($\mu_{+}(0)$) is not immediately known.  The right-going states
of the two layers move toward equilibrium moving from left to
right along the edge in accord with Eq.~(\ref{chemsolv}), reaching
$\mu_{-}(L)$ and $\mu_{+}(L)$ as the drain contact is approached.
Similarly along the left-going edge the chemical potential of the contacted
layer at the right-side of the sample is fixed by the
chemical potential of the drain ($\mu_{D}$) and the chemical
potential of the floating layer is $\mu_{+}(L)$.  The
equilibration of the two-layers along the left going edge is
again described by Eq.~(\ref{chemsolv}).  Setting the value of the
effective chemical potential at the left side of the floating layer
to $\mu_{+}(0)$
gives a self-consistent equation which allows the relative
values of all effective
chemical potentials to be related to the net current flowing
through the sample.  We find that
\begin{equation}
\mu_{S} - \mu_{D} = \frac{hI}{e}[ \frac{1}{1+\exp(-2L/\xi)}+\frac{1}{2}],
\label{msmd}
\end{equation}
\begin{equation}
\mu_{+}(0) = \frac{hI}{2e} + \mu_{D},
\label{ms}
\end{equation}
where $L$ is the distance along the edges between
source and drain.   Because the equilibration between
contacted and floating layers occurs gradually along the layer,
the measured longitudinal ($R_{xx}=R_{12} = R_{34}$)
and Hall ($R_{xy}=R_{13}$) resistances are dependent on the
placement of the voltage probes:
\begin{equation}
R_{xx} = \frac{h}{e^{2}}
\frac{\exp(-2a/\xi)[1 - \exp(-2b/\xi)]}{2[1+\exp(-2L/\xi)]},
\label{rxx}
\end{equation}
\begin{equation}
R_{xy} = \frac{h}{e^{2}}
\frac{1+\exp(-2a/\xi)+\exp[-2(L-a)/\xi] + \exp(-2L/\xi)}
{2[1+\exp(-2L/\xi)]}.
\label{rxy}
\end{equation}
where $a$ and $b$ specify the voltage probe
positions as indicated in Fig.~1.

Note that in the limit of weak inter-layer coupling ($\xi \to \infty)$,
$R_{xx}$ which would be zero for decoupled layers has a correction at
first order in $\xi^{-1}$: $R_{xx} \simeq
(h/e^{2})(b/2\xi)$.  In a Landauer-Buttiker picture
dissipation is associated with back-scattering between left-going
states on one sample edge and right-going states on the
opposite sample edge.  In this system backscattering in the contacted
layer occurs indirectly by tunneling to the
floating layer and drifting around its perimeter to the other side
of the sample before tunneling back to the contacted layer.
The weak-coupling correction to the quantized Hall resistivity
value, ($h/e^{2}$), is second order in $\xi^{-1}$ and vanishes as
the voltage probes go toward the end of the Hall bar:
$R_{xy} \simeq (h/e^{2})[1 - a(L-a)/\xi^{2}]$.
Thus $R_{xy}$ can be nearly quantized even when $R_{xx}$ is
a substantial fraction of $h/e^{2}$.
In the opposite limit of strong inter-layer coupling($\xi \to 0$),
$R_{xx}$ vanishes, and $R_{xy}$ takes a quantized value of
$h/2e^{2}$.
In this strong coupling limit we get the same results as we would
for a system where both layers are contacted.
The overall behavior of resistances are shown in Fig.~2 for
the choice of $a/L= 0.2$ and $b/L= 0.6$.
The non-locality of the resistances, which depend on $a$, $b$, and
$L$ should be noted.
In general the peak of $R_{xx}$ occurs around $L \simeq \xi$
for a symmetrical system where $L=2a+b$.
The approximate value of the peak is $(b/3L)(h/e^{2})$, except when
$b$ is quite close to $L$ in which case it increases more rapidly and
reaches the limiting value $h/2e^{2}$ at $b=L$.
\par
The present theoretical scheme can be applied to various
other lead geometries, including ones where leads are also attached
to the floating layer.  If only voltage probes are attached to this layer,
it is possible to measure the ``transresistances",
$R_{xx}^{t}$ and $R_{xy}^{t}$, the ratios
of voltages measured in the open layer to the net
current \cite{eisenexp,pepperexp,sivanexp}.
Our scheme gives $R_{xx}^{t} = - R_{xx}$,
and $R_{xy}^{t} = h/e^{2} - R_{xy}$ in this case.
Current leads can also be attached to the both layers.
Experimentally it is easy to let either the source or the
drain make contact to both layers.
The resistivities measured in the lower layer
for the case where the source is connected to both layers and the
drain is connected only to the lower layer is
$R_{12} = 0$,
$R_{13} = ({h}/{2e^{2}}) \{1 + \exp[-2(a+b)/\xi]\}$,
and
$R_{34} = ({h}/{2e^{2}})\exp(-2a/{\xi})
[1 - \exp({-2b}/{\xi})],$
Thus in this case the correction to the quantized Hall resistance is
also linear in $\xi^{-1}$ \cite{sourcedrain}.
Another geometry of interest is realized when the voltage probes in the top
layer are used as current leads.
When the two current leads to the top layer are connected by
an ideal conductor, current through the leads, $I_{+}$,
is induced by the current in the minus layer, $I_{-}$.
For simple situations where $a$ in Fig.~1 is 0 or $L/2$,
the current through the leads 1 and 4 is given by
$I_{+,14}/I_{-} = [1-\exp(-2L/\xi)]/[3+\exp(-2L/\xi)]$,
and
$I_{+,14}/I_{-} = [1-\exp(-L/\xi)]^{2}/[3+\exp(-2L/\xi)]$,
respectively,
where $I_{-}$ is the current through the source and drain in the minus layer,
as usual.
In both cases, the ratio of the currents can be as large as 1/3
in the strong coupling regime.
We remark that the transresistances
and induced currents discussed here originate
purely from tunneling between the two layers.
Similar effects can also be provided by
the frictional drag by the Coulomb interaction
\cite{eisenexp,martin} although the drag resistances
tend to be much smaller.
\par
%
%
We now compare these results with the experiments
by Ohno and Sakaki \cite{ohno,ohno1}.
Their results, Fig.~3 in ref.~\cite{ohno} clearly show
that the experimental situation is in the weak inter-layer tunneling regime.
At first sight this is surprising since the hopping amplitude
in the bulk of their samples
, responsible for the splitting between symmetric
and antisymmetric subbands, was estimated to be
$\Delta_{{\rm SAS}} \simeq 0.02 {\rm meV}$.
If the edges of their two samples were perfectly aligned we
would estimate that $\xi = \hbar v_{F} /\Delta_{\rm SAS}$ where
$v_{F} \sim \ell \omega_{c}$ is the velocity of edge states in the abrupt edge
limit.   ($\ell$ is the magnetic length
and $\omega_{c}$ is the cyclotron frequency.)  Since $L = 200\mu$m, this
estimate gives $L/\xi \sim 50$, well into the strong coupling
regime.   A likely cause of this discrepancy
is disorder at the edge which results in random misalignment of the edge
states in the two layers.
Such a situation can be caused by
roughness at the mesa-etched sample wall.
Since vertical tunneling is possible only where the edges
are aligned, disorder will
drastically increase the tunneling length \cite{dy}.
Support for this picture is found in the observed \cite{ohno1}
 weak dependence of coupling on the orientation of the magnetic field with
respect to the normal to the two-dimensional layers.
Even when the magnetic field is tilted in the direction perpendicular
to the current direction, the values of the resistance shows little
change.  If the edges were perfectly aligned, in-plane magnetic field
would reduce the hopping amplitude,
and reduce the minimum value of $R_{xx}$ noticeably.
However, such a reduction does not occur when the hopping amplitude has
already been reduced by the misalignment \cite{dy}.
\par
A remarkable result of the experiment is that there are peaks in
$R_{xx}$ on the high field side of the QH plateaus \cite{ohno}.
These peaks occur between the $R_{xx}$ minimum and the ordinary
peaks of $R_{xx}$ which exist between the plateaus.
This peak in the higher field side of the plateau
can be understood as a phenomenon caused by the increase of the
hopping amplitude as the magnetic field is increased:
On the high field side,
the Fermi edges continuously move towards the center of the
sample.
This causes three effects:
(i) $v_{F}$ is reduced,
(ii) the size of the edge region
increases,
and (iii) the alignment of the two edges improves.
All of these effects make $L/\xi$ larger.
Figure 2 tells us that if the hopping amplitude is continuously
increased from near the origin,
$R_{xy}$ decreases from $h/e^{2}$ towards $h/2e^{2}$,
and $R_{xx}$ shows a peak.
It should be noted that in our model the height of the peak
depends only on the geometry of the sample,
and on the ratios $a/L$ and $b/L$.
Our estimates for the height, about 5k$\Omega$ (2.5k$\Omega$) for
Landau level filling factor 1 (2), are consistent with the
experiment \cite{ohno,ohno1,note2}.
Our theory suggests that this additional peak in $R_{xx}$ and
the nearly quantized Hall plateau occur in
systems studied in these experiments because their bulk states are strongly
coupled but their edge states are close to the weak coupling limit.
The experimental results also show peaks
on the low field side of the $R_{xx}$ minimum.
This peak is not separated from the ordinary peak of $R_{xx}$.
This peak originates from the bottom of the higher Landau level,
so that conduction in the bulk is important and our theory cannot
be applied.
\par
%
%

In summary we have developed a theory which relates the transport properties
of double-layer 2DES's near quantum Hall plateaus to inter-layer
equilibration lengths along the sample edges.
Our results permit a consistent interpretation of
recent floating-layer experiments
and can be applied to other contacting
geometries.  We anticipate that our theory will be useful
in interpreting experiments whose aim is to probe the disorder and
interaction physics of quantum Hall effect edge states.
\par
%
%
We thank Yuzo Ohno for sharing his experimental results
prior to the publication and Steve Girvin for helpful
comments.   This work was supported by the National Science Foundation
under grant DMR-9416906.
\par
%
%

%
%
\begin{figure}
\caption{Geometry of the sample.  
Two identical layers are stacked vertically.  The leads
(source, drain, 1, 2, 3, and 4) can contact both layers or either layer.
The drift directions along the edges are indicated by arrows.
The lengths $L$, $a$, and $b$ specify the length of the sample and
the positions of the leads.}
\label{fig1}
\end{figure}
\begin{figure}
\caption{Longitudinal and Hall Resistances as a function of
$L/\xi$.
Here $\xi$ is the inter edge relaxation length.}
\label{fig2}
\end{figure}


\begin{references}
\bibitem{note1}  The temperature drops
out in this equation.  As long as $kT \ll \hbar\omega_c$
our results are tempreature independent.
\bibitem{sourcedrain} It should be noted the definition of
the source and drain is relative
to leads and edge current directions.
Therefore, for example, if the magnetic field is reversed,
the role of the source and the drain should be interchanged.
\bibitem{note2} If $\xi$ is not constant along the edge, the height
of the peak will change.  In the case of filling factor larger
than one the peak height can also be slightly different,
since the maxima from the different Landau levels need not be
realized at the same time.
\end{references}
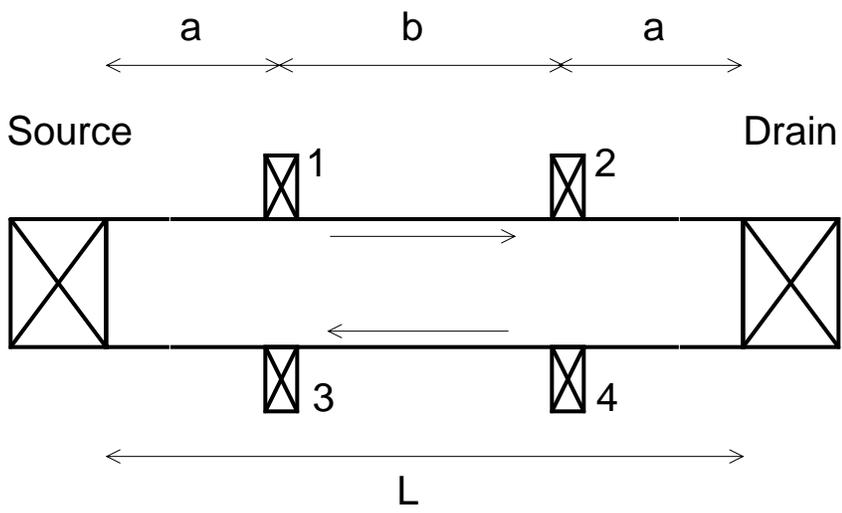
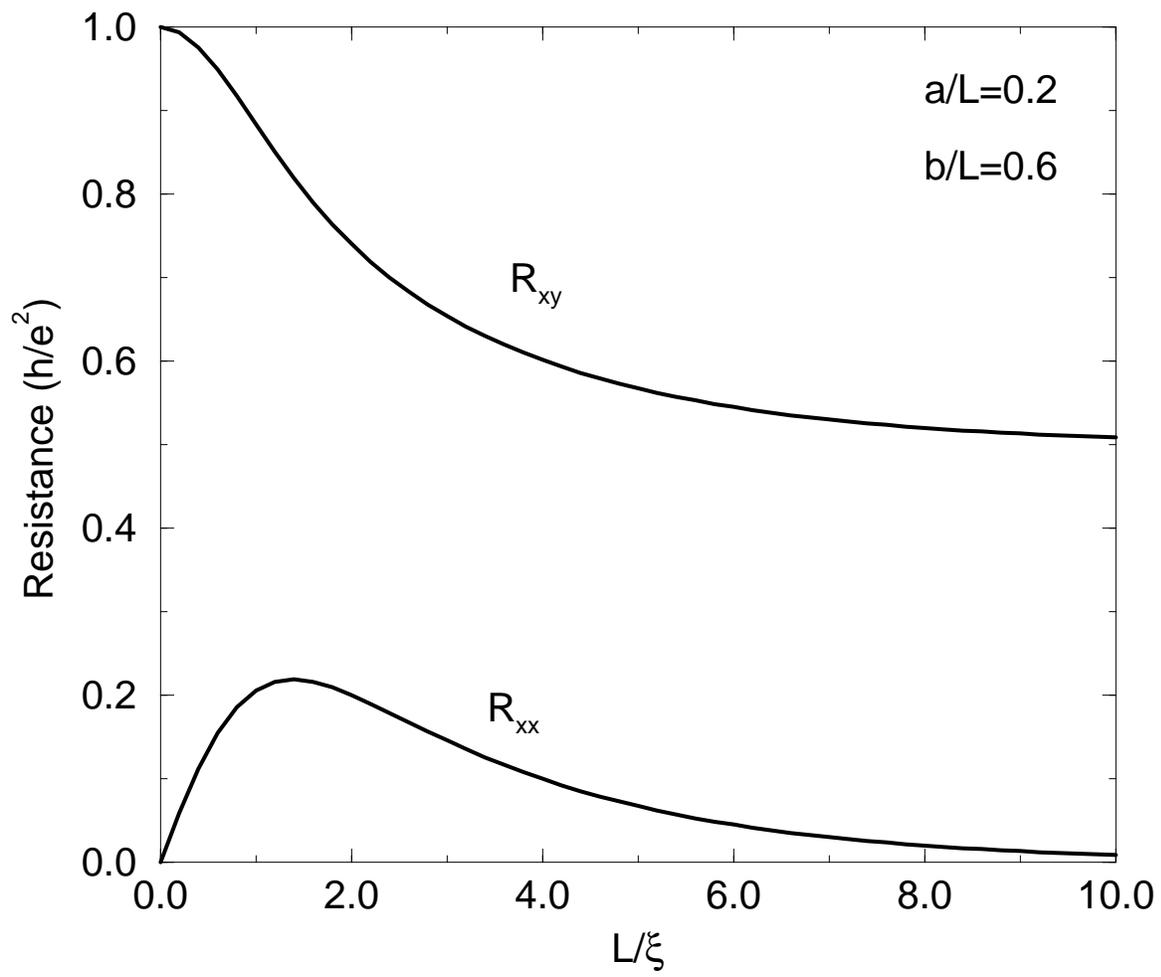
\end{document}